\documentclass[aps,preprint,showpacs]{revtex4}
\usepackage{amsmath}
\usepackage{color}
\usepackage{graphicx}

\nocite{*}
\begin{document}

\title{Birefringence and QuasiNormal Modes of the Einstein-Euler-Heisenberg Black Hole}

\author{Nora Bret\'on}
\email{nora@fis.cinvestav.mx}
\affiliation{Departmento de F\'{\i}sica, CINVESTAV--IPN, Apartado Postal 14--740, C.P. 07000, M\'exico City, M\'exico.}
\author{L. A. L\'opez}
\email{lalopez@uaeh.edu.mx}
\affiliation{\'Area Acad\'emica de Matem\'aticas y F\'isica, UAEH, Carretera Pachuca-Tulancingo Km. 4.5, C P. 42184, Mineral de la Reforma, Hidalgo, M\'exico.}
\date{\today}

\begin{abstract}
In this contribution we study the birefringence  and the quasinormal modes (QNM) in the eikonal approximation, 
of the Einstein-Euler-Heisenberg black hole (EEH-BH) . The EEH-BH is an exact solution of the Einstein equations coupled to the Euler-Heisenberg nonlinear electrodynamics. It is well known that in the presence of strong electromagnetic fields  light rays modify their trajectories and  do follow the null geodesics of an effective metric as a result of the nonlinear interaction between light and the intense magnetic or electric background. In the Euler-Heisenberg theory the phenomenon of birefringence arises and then exist  two possible light trajectories for polarized light in the vicinity of the EEH-BH. On the other hand, using the correspondence between the parameters of the null unstable geodesics and the quasinormal modes in the eikonal approximation. We determine these QNM for both, the electric and the magnetic EEH-BH and  compare them with its linear counterpart, the Reissner-Nordstr\"{o}m black hole, since there are two effective metrics, to each one corresponds one null geodesic that renders two possible QNM from the same compact object. 
\end{abstract}


\maketitle

\section{Introduction}

These days it has been widely accepted the existence of black holes (BH), or at least of very compact astrophysical  objects that resemble several characteristics ascribable to a BH.
In most of the observations BH are accompanied by magnetic fields of varied intensities, that go from $10^{-6}$ Gauss in the centers of galaxies to orders of $10^{12}$ Gauss for neutron stars  \cite{Gold1968}. These magnetic fields are of unknown origin, so far; in neutron stars they may have an internal origin. In case of strong magnetic fields they very likely produce vacuum polarization in their vicinity, then processes like photon splitting, pair conversion or vacuum polarization are expected to occur in the neighborhood of neutron stars \cite{Baring2008}.


In the presence of intense electromagnetic fields quantum electrodynamics (QED) predicts that vacuum has properties of a material medium as a consequence of the electromagnetic  self-interactions. These effects become significant when  electromagnetic fields approach the critical strengths, $E_c \approx 10^{18}$ Volt/m or $B_c \approx 10^{13}$ Gauss; among them are light-light interaction or electron-positron creation (vacuum polarization). Therefore 
an electromagnetic (EM) wave traveling through intense EM fields will change its velocity and the direction of propagation may depend on its polarization, this later effect is the birefringence  \cite{Bialynicka1970}, \cite{Brezin1971}, \cite{deMelo2015}.
Experimental efforts are currently in progress to observe in laboratory these nonlinear electrodynamics (NLED) effects; we mention just a few of them, like detection of vacuum birefringence with intense laser pulses \cite{Luiten2004}, \cite{Gies2009}, \cite{Karbstein2020}, or using waveguides \cite{Brodin2001}.
 
 By treating the vacuum as a medium, the Euler-Heisenberg (EH) theory  \cite{EK1935}, \cite{EH1936}, predicts rates of nonlinear light interaction processes since
it takes into account vacuum polarization to one loop, and is valid for electromagnetic fields that change slowly compared to the inverse electron mass. The EH Lagrangian depends in nonlinear way of the two Lorentz and gauge invariants, $F$ and $G$,

\begin{equation}
\mathcal{L}_{\rm EH} (F,G) = -\frac{F}{4}+ \frac{\mu}{4}\left(F^{2}+\frac{7}{4}G^{2} \right),
\label{E-HLagrangian}
\end{equation}
where  $F=F^{\mu \lambda}F_{\mu \lambda}= 2(B^2-E^2)$ and $G=-{}^{\ast}F^{\mu\lambda}F_{\mu\lambda}= 4 \vec{B} \cdot \vec{E}$, $F_{\mu \lambda}$ being the Faraday tensor, and   $^*F_{\mu\nu}=\frac{1}{2\sqrt{-g}}\epsilon_{\mu\nu\rho\sigma}F^{\sigma\rho}$  its dual, while $\mu$ is the parameter of the EH theory that in terms of the fine structure constant, $\alpha$, is
\begin{equation}
\mu = \frac{2 \alpha^2}{45m_e^4};
\label{b}
\end{equation}
or in terms of the critical fields, is of the order $\mu \sim \alpha/ B_c^2$. 
The linear electromagnetic Maxwell theory is recovered if $\mu=0$, then $\mathcal{L}_{\rm Maxwell} (F) = -{F}/{4}$. 
The Lagrangian in (\ref{E-HLagrangian}) is actually the Euler-Kockel Lagrangian \cite{EK1935}, that is the EH Lagrangian expanded up to second order in $\alpha$. 
Coupling the EH theory with gravitation BH solutions can be obtained;  the static spherically symmetric solution represents a BH with an electric (or magnetic) charge that generates such intense EM fields in which NLED effects arise like birefringence of light rays.

On the other hand,  a BH may be characterized by relating its independent parameters to the quasinormal modes (QNM) that result from the 
response of a test field perturbing the BH spacetime. QNM has been one of the most useful tools on BH characterization. 
In the geometric-optics or eikonal approximation QNM can be determined from the unstable null geodesics that are the orbits attached to the maximum of the effective potential barrier felt by light rays on their interaction with the BH.
In 1984 Ferrari and Mashhoon \cite{Ferrari1984} suggested an analytical technique of calculating the QNM in the eikonal limit and 
later on Cardoso \cite{Cardoso2008} showed  the relationship among unstable null geodesics, Lyapunov exponents and QNM for a stationary spherically symmetric space-time.  The idea basically  consists in  interpreting the black-hole free oscillations in terms of null particles trapped at the unstable circular orbit and slowly leaking out. The real part of the complex QNM frequencies is determined by the angular velocity at the unstable null geodesic; the imaginary part is related to the instability time scale of the orbit.
As compared with the WKB method calculating the QNM in the eikonal approximation turns out to be a good one for large angular momentum perturbations in the lowest modes.

Unstable null geodesics are derived from the metric; however,
as a result of the nonlinear interaction, light rays do not follow the null geodesics of the background metric, but do follow the null geodesics of an effective metric that depends on the nonlinear electromagnetic energy momentum tensor \cite{Pleban}, \cite{Novello2000}. The EEH-BH is characterized by a strong field in  its vicinity, an EM field of such strength  produces birefringence, i.e. two different trajectories for light rays, that are the null geodesics of the effective metrics derived from the study of the propagation of the characteristic surfaces of the EM field \cite{Pleban}, \cite{Novello2000}.


In this contribution we determine the two effective metrics whose null geodesics are the light trajectories as well as the QNM in the eikonal approximation arising from perturbing the static EEH-BH.
Due to the birefringence effect  QNM of two different frequencies will arise, 
corresponding  to the two different effective metrics emerging due to the NLED interaction of light rays and the strong field background. We determine the unstable null circular orbits that are followed by light rays and then the QNM in the eikonal approximation. We address both, electric and magnetic EEH-BH.

The paper is organized as follows:
In the next Section a short
summary of the EH NLED and the static spherically symmetric solution of the EEH equations are presented.
In Section III, we derive the two effective metrics that arise in the EH NLED for both cases, electric and magnetic charge.
In  Section IV we give a brief explanation for determining  the QNM using the unstable null geodesic and the Lyapunov exponent, as well as
the expressions in terms of the effective potential and the radios of the null unstable circular orbits. In Section V we present the explicit expressions for the QNM of the EEH-BH and the linear limit, the Reissner-Nordstr\"{o}m(RN) BH. In the same section we analyze  the QNM for the  EEH-BH, and compare them with  the ones from massless test particles (light rays) in the RN-BH.  Final remarks are given in the  Section VI.

\section{The Einstein-Euler-Heisenberg Black Hole }

The four dimensional action of general relativity coupled to the EH nonlinear electrodynamics (NLED)  with Lagrangian
$\mathcal{L}_{\rm EH} (F,G) $ is 

\begin{equation}
S=\frac{1}{4\pi}\int_{M^4} d^4x \sqrt{-g}\left[\frac{R}{4}-
\mathcal{L}_{\rm EH} (F,G)  \right],\label{action}
\end{equation} 
where $g$ is the determinant of the metric tensor, $R$ is the Ricci scalar and $\mathcal{L}_{\rm EH} (F,G) $
is the  EH Lagrangian in Eq. (\ref{E-HLagrangian}).

Regarding NLED
there are two possible frameworks \cite{Pleban}, one of them is the usual one ($F$-framework) in terms of
the electromagnetic field tensor $F^{\mu \nu}$.
Alternatively, there is the $P$-framework with the tensor $P_{\mu\nu}$ as the main field,
defined by
\begin{equation} 
P_{\mu\nu}= -(\mathcal{L}_F F_{\mu\nu}+ {^*F}_{\mu\nu} \mathcal{L}_G ),
\label{Pmunu}
\end{equation}
where the subscript $X$ in $\mathcal{L}$ denotes the derivative, $  \mathcal{L}_{X}= d
\mathcal{L} /d X$. In  the Euler-Heisenberg theory, $P_{\mu \nu}$ takes the form
\begin{equation}
P_{\mu\nu}=(1-\mu  F)F_{\mu\nu} - {^*F}_{\mu\nu} \frac{7 \mu}{4}G .\label{Pmunu_a1}
\end{equation}
The tensor $P_{\mu \nu}$ corresponds to the electric displacement {\bf D} and the
magnetic field {\bf H}  while $F_{\mu \nu}$  corresponds to the magnetic intensity {\bf B} and the electric field {\bf E},
and  Eqs. (\ref{Pmunu_a1}) are the constitutive relations
between  ({\bf D},  {\bf H}) and ({\bf E},  {\bf B})  in the EH NLED. \\

The two NLED frameworks, $F$ and $P$, correspond to the  Lagrangian and Hamiltonian treatments, respectively.
The two invariants $s$ and $t$ associated to the P framework are defined as
\begin{equation}
s=-\frac{1}{4} P_{\mu\nu} P^{\mu\nu}, \hspace{1cm}
t= -\frac{1}{4} P_{\mu\nu} {^*P^{\mu\nu}},
\label{Invst}
\end{equation}
with $^*P_{\mu\nu}=\frac{1}{2\sqrt{-g}}\epsilon_{\mu\nu\rho\sigma}P^{\sigma\rho}$ the dual tensor to $P_{\mu \nu}$.
The Legendre transformation of $\mathcal{L}$ defines the Hamiltonian or structural
function $\mathcal{H}$
\begin{equation}
\mathcal{H} (s,t)= -\frac{1}{2}P^{\mu\nu} F_{\mu\nu}-\mathcal{L}.
\end{equation}
Neglecting  the second and higher order terms in $\mu$, the structural function for the EH
theory takes the form \cite{Remo2013}
\begin{equation} 
\mathcal{H}(s,t)= s- 4\mu \left( s^2 + \frac{7\mu}{4}t^2 \right). \label{HamiltonianEH}
\end{equation}

The EM and gravitational  field equations are then
\begin{equation}
\nabla_\mu P^{\mu\nu}=0, \hspace{.7cm} G_{\mu\nu}+\Lambda g_{\mu\nu} =8\pi
T_{\mu\nu}.\label{motion}
\end{equation}
The energy momentum tensor $T_{\mu\nu}$ for the EH theory in the $P$ framework is given by

\begin{equation}
T_{\mu\nu}=\frac{1}{4\pi}\left\{ (1- \mu s)P_\mu^\beta
P_{\nu\beta}+g_{\mu\nu}\left(s -\frac{\mu }{2} \left[
3 s^2 + \frac{7}{4}t^2 \right] \right)\right\}.\label{emtensor}
\end{equation}

In the next subsections we present the static spherically symmetric solution of the EEH equations, in both cases, electric and magnetically charged.

\subsection{Electrically charged EEH--BH solution}

The solution to  Eqs. (\ref{motion}) for a static, spherically symmetric (SSS) metric of the form
\begin{equation}
ds^2= -f(r)dt^2 + f(r)^{-1}dr^2 + r^2(d\theta^2 +\sin^2{\theta} d\varphi^2),
\label{metric}
\end{equation}
with $f(r)=1-{2m(r)}/{r}$,   was derived in \cite{Amaro2020} using the NLED P-framework \cite{Pleban}.
The metric function for the electric case is given by

\begin{equation}
f(r)= 1-\frac{2M}{r}+\frac{Q_e^2}{r^2}-\frac{\mu Q_e^4}{20 r^6},
\label{gtt}
\end{equation}
where $M$ is the mass of the BH, $Q_e$ its electric charge and $\mu$ is the EH parameter.
In (\ref{gtt}) the case $\mu =0$ corresponds to the Reissner-Nordstr\"{o}m (RN) solution, that is the SSS solution to the Einstein-Maxwell equations.
Then the last term is the extra one compared to the RN metric function,  and it can be considered as a kind of screening of the BH charge due to the vacuum polarization effect \cite{Remo2013, Amaro2020}; 
the screening effect is clear writing the metric as,

\begin{equation}
f(r)= 1-\frac{2M}{r}+\frac{Q_e^2}{r^2}  \left\{ 1 -\frac{\mu Q_e^2}{20 r^4} \right\}.
\label{gtt1}
\end{equation}
In general the behavior of $f(r)$ is Schwarzschild-like in contrast  with other NLED BHs that have a RN behavior but with a screened charge (for instance Born-Infeld BH \cite{Mann}, \cite{Kruglov2010}). The singularity remains at $r=0$
and is stronger and of opposite sign than in RN. The equation that determines the horizons $f(r_+)=0$ is a six degree polynomial,

\begin{equation}\label{HNo}
r_+^6  - 2 r_+^5 + Q_e^2 r_+^4 -  \frac{\mu Q_e^4 }{20} =0,    
\end{equation}
the previous Eq. has been written in terms of $r \mapsto r/M$ and the  dimensionless parameters, $Q \mapsto Q/M$ and $\mu \mapsto \mu/M^2$.
The number of horizons may vary from three to one. When we apply the method described in \cite{Toshmatov2017}, we determine the range of values for $Q_e$ and $\mu$, so that the line element of Euler--Heisenberg in the electric case (\ref{gtt}), represents a black hole or an extreme black hole. For $\mu/M^{2} \leq 50/81$ and $Q_e^{2}/M^{2}\leq 25/24$ the equation (\ref{HNo}) may have three real positive roots (one outer horizon and two inner horizons); the extreme case can be obtained from the conditions $f(r) = 0$ and $df(r)/dr =0$, that amounts to the EH parameter $\mu$ being

\begin{equation}
\mu_{\rm ext}=\frac{5^{5}\left(1 \pm\sqrt{1-24Q_e^{2}/25}\right)^{4}\left(6Q_e^{2}-5 (1 \mp  \sqrt{1-24Q_e^{2}/25})\right)}{18^{3} Q_e^4}.
\end{equation}
 
In the P-framework the electromagnetic field is given by the anti-symmetric tensor $P_{\mu \nu}$ that for a SSS metric has the form 

\begin{equation}
P_{\mu\nu}= \frac{Q_e}{r^2}(\delta^1_{\mu} \delta^0_{\nu}- \delta^0_{\mu} \delta^1_{\nu} ); \label{Pmunu01}
\end{equation}
then the EM invariants, $s$ and $t$ in Eq. (\ref{Invst}), are
\begin{equation}
s=\frac{Q_e^2}{2r^4},\hspace{.7cm} t=0  \label{st} . 
\end{equation}  

The tensor
$P_{\mu \nu}$ is related to the Faraday 
tensor $F_{\mu \nu}$ by the constitutive or material relations,

\begin{equation}
F_{\mu \nu}= \left( {1- \mu s - \frac{ 7 \mu}{4} t} \right)    P_{\mu \nu};
\end{equation}
therefore in this case the nonvanishing component of the Faraday tensor is

\begin{equation}
F_{01}= \left(  1-  \frac{\mu  P_{01}^2}{2} \right) P_{01}.
\label{emfield}
\end{equation}
\subsection{Magnetically charged EEH--BH solution}

The SSS solution of the EEH magnetically charged has the same metric component $g_{tt}$ in Eq. (\ref{gtt}) but replacing the electric charge with a magnetic one, $Q_e \mapsto Q_m$. The magnetic case is more conveniently obtained in the F-framework of NLED; the magnetic field is given by

\begin{equation}
F_{\mu \nu} =    Q_m \sin{\theta}(\delta^3_{\mu} \delta^2_{\nu}- \delta^2_{\mu} \delta^3_{\nu} ),
\label{magnF}
\end{equation}
while the invariants in this case are $F=2 Q_m^2/r^4$ and $G=0$.
The magnetic EEH--BH horizons also have been analyzed in \cite{Yajima2001}.
In \cite{Rubiera2019} is treated the electric case in the  NLED $F$-framework;
see also \cite{Kruglov2017}. Recently, from the electric static EEH solution 
has been derived a stationary EEH solution \cite{Macias2019}.


\section{Effective metrics of the EEH--BH}

The theory defined by the  Lagrangian $\mathcal{L}_{\rm EH} (F,G) $  (\ref{E-HLagrangian}) admits the phenomenon of birefringence, by this meaning that light rays with different polarization do follow distinct trajectories.  These trajectories are determined by the null geodesics of an effective metric  (or  pseudometric \cite{Pleban}), that depends on the  matter tensor. The effective metric can be calculated from the study of the characteristic surfaces or the propagation of discontinuities of the electromagnetic field \cite{Pleban}.  In this treatment, that is equivalent to the {\it soft photon approximation} \cite{Novello2000b}, \cite{Liberati-Sonego-Visser}, a system of coupled equations is derived in \cite{Novello2000},
that for the EH case, in which $\mathcal{L}_{F G}=0$, can be decoupled into two effective metrics, $\gamma^{(i) \mu \nu}, \quad i=1,2,$ given by

\begin{eqnarray}
\gamma^{(1) \mu \nu} & = & \left(\mathcal{L}_{F }- 2 \mathcal{L}_{G G} F \right) g^{\mu \nu} - 
4 \mathcal{L}_{G G} F^{\mu}_{. \lambda} F^{\lambda \nu}, \nonumber\\
\gamma^{(2) \mu \nu} & = & \mathcal{L}_{F } g^{\mu \nu} -  4 \mathcal{L}_{F F} F^{\mu}_{. \lambda} F^{\lambda \nu}, 
\label{effmetrics}
\end{eqnarray}
where $g^{\mu \nu}$ is the background metric, in our case is the EEH--BH metric, Eq. (\ref{metric}) with (\ref{gtt}).
See also \cite{Obukov2002}, \cite{Goulart2009}.
Another NLED that exhibit birefringence was  explored in \cite{Kruglov2015}.
\subsection{Effective metrics  for the electric EEH--BH}
 
Given a SSS metric $g_{\mu \nu}$, of the form Eq. (\ref{metric}), and the electromagnetic field in Eq. (\ref{emfield}), explicitly 
the two effective metrics of Eqs. (\ref{effmetrics}) become

\begin{eqnarray}
\gamma^{(1) \mu \nu} & = & \left( - \frac{1}{4} + \frac{5}{2} \mu P_{01}^2 \right) g^{\mu \nu} - 
\frac{7}{2} \mu  F^{\mu}_{. \lambda} F^{\lambda \nu},
\label{effmetr1}\\
\gamma^{(2) \mu \nu} & = & \left( - \frac{1}{4} - \mu P_{01}^2 \right) g^{\mu \nu} -  2 \mu  F^{\mu}_{. \lambda} F^{\lambda \nu}, 
\label{effmetr2}
\end{eqnarray}
where we have substituted the EM invariant $F$ given by

\begin{equation}
F= - 2 F_{01}^2 = -2 (1- a P_{01}^2) P_{01}^2,   
\label{F}
\end{equation}
and  we have neglected terms of $\mathcal{O} (\mu^2)$, to be consistent with the EH Lagrangian that is valid up to $\mu$--order.
The effective metrics (\ref{effmetr1}) and (\ref{effmetr2}), up to a conformal factor that leave null geodesics invariant, can be comprised in the formula for the line element,

\begin{equation}
\gamma^{(i)}_{\mu \nu} dx^{\mu} dx^{\nu}= \frac{1}{G^{e}_{i}(r)} \left( -f(r) dt^2 + \frac{dr^2}{f(r)} \right) + r^2 d\Omega^2, \quad i=1,2,   
\end{equation}
with

\begin{equation}
G^{e}_{1}(r) = 1 - \left(   \frac{4 \mathcal{L}_{GG} (F_{10})^2}{\mathcal{L}_F - 2 \mathcal{L}_{GG} F} \right) =
1+ \mu \frac{14 (F_{01})^2}{1 + 5 \mu F},  \quad
G^{e}_{2}(r) = 1 - \left( \frac{4 \mathcal{L}_{FF} (F_{10})^2}{\mathcal{L}_F} \right)= 1+ \mu \frac{8 (F_{01})^2}{1 - 2 \mu F},  
\end{equation}
The factors $G^{e}_i$ become 1 in the linear case, and there is no biregringence effect.
Notice that the terms responsible for birefringence  depend on $\mathcal{L}_{GG}$ or $\mathcal{L}_{FF}$, therefore, birefringence arises only from NLED Lagrangians depending on the EM invariants in nonlinear way, otherwise if $\mathcal{L}_{GG}=0$ or
$\mathcal{L}_{FF}=0$, there is not birefringence.
In the previous expressions we must keep terms up to first order in $\mu$ (second order in $\alpha$), obtaining then

\begin{equation}
G^{e}_{1}(r) = 1+ 14 \mu \frac{Q_e^2}{r^4},  \quad
G^{e}_{2}(r) = 1+ 8 \mu \frac{Q_e^2}{r^4},
\end{equation}
\subsection{Effective metrics  for the magnetic EEH--BH}

For the magnetic case, the electromagnetic field is given by Eq. (\ref{magnF}), and there is as well birefringence, 
the effective metrics in Eqs. (\ref{effmetrics}) now are given by

\begin{equation}
\gamma^{(i)}_{\mu \nu} dx^{\mu} dx^{\nu}= {G^{m}_{i}(r)} \left( -f(r) dt^2 + \frac{dr^2}{f(r)} \right) + r^2 d\Omega^2, \quad i=1,2, \end{equation}
with

\begin{equation}
G^{m}_{1}(r) = 1 - 12 \mu \frac{Q_m^2}{r^4}, \quad
G^{m}_{2}(r) = 1 - 4 \mu \frac{Q_m^2}{r^4},  
\end{equation}
where $Q_m$ is the magnetic charge and we only kept terms up to $\mathcal{O}(\mu)$.
\section{QNM and  Unstable Null Geodesics in SSS spaces}

The connection between the QNM and bound states of the the inverted black hole effective potential was pointed out in 
\cite{Ferrari1984}. 
In \cite{Cardoso2008}  it was shown that, in the eikonal limit,  the QNMs of black holes in any dimension are determined by the parameters of the circular null geodesics. The real part ($\omega_{r}$) of the complex QNM frequencies is determined by the angular velocity at the unstable null geodesics, while the imaginary part ($\omega_{\rm im}$), that is related to the instability time scale of the orbit, is related to the Lyapunov principal exponent. In the case of stationary, spherically symmetric spacetimes  it turns out that this exponent can be expressed as the second derivative of the effective potential evaluated at the radius of the unstable circular null orbit, in such a way that the QNM are given by

\begin{equation}\label{QNM}
\omega_{QNM}=\Omega_{c}l-\imath (n+\frac{1}{2})| \lambda|,
\end{equation}
where $n$ is the overtone number and $l$ is the angular momentum of the perturbation. $\Omega_{c}$ is the angular velocity at the unstable null geodesic and  $\lambda$ is the Lyapunov exponent, determining the instability time scale of the orbit. The Lyapunov exponents are a measurement  of the average rate at which nearby trajectories converge or diverge in the phase-space. A positive Lyapunov exponent indicates a divergence between nearby trajectories, i.e., a high sensitivity to initial conditions.  From the equations of motion and using the definition $\dot{r}^{2} + V(r)=0$, where $V(r)$  is the effective potential for radial motion, circular geodesics are determined from the conditions $V(r_{c})=V^{'}(r_{c})=0$ where $r_c$ is the radius of the circular orbit.
The  Lyapunov exponent in terms of the second derivative of the effective potential; is given by

\begin{equation}\label{expLyapunov}
\lambda=\sqrt{\frac{-V^{''}}{2\dot{t}^{2}}},
\end{equation}

where $t$ is the time coordinate. The dot denotes the derivative with respect to an affine parameter of the geodesic while the  prime stands for the derivative with respect to $r$ . 

The orbital angular velocity is given by

\begin{equation}\label{angular}
\Omega_{c}=\frac{d{\varphi}}{d{t}}=\frac{\dot{\varphi}}{\dot{t}}.
\end{equation}

For our purpose both expressions should be evaluated at $r_c$, the radius of the  unstable null circular orbit, denoted by the $c$ subscript; that is the orbit with an impact parameter $b=L/E$ and with $V^{''} (r_c) < 0$.  

For a static  spherically symmetric (SSS) metrics of the form Eq. (\ref{metric}), the energy  $E$   and the angular momentum $L$ of a test particle  are conserved quantities,

\begin{equation}\label{constan}
f(r)\dot{t}=E= {\rm const},\;\;\;\;\;\;\;\ r^{2}\dot{\varphi}=L= {\rm const}.
\end{equation}

For equatorial orbits, from the equation of radial motion, $\dot{r}^{2} + V(r)=0$, in the static space-time,  the effective potential is given by

\begin{equation}
V(r)= E^2 \left(1-  \frac{f(r)}{r^2} \frac{L^{2}}{E^{2}} \right).
\end{equation}

Then the Lyapunov exponent, related to the imaginary part of the QNM,  from (\ref{expLyapunov}) is given by

\begin{equation}\label{Lyapunov}
\lambda^{2}=\frac{f}{2r^{2}}\left[2f-r^{2}f^{''}\right]\bigg |_{r_c},
\end{equation}

while the orbital angular velocity, that is proportional to the QNM real part,  is given by

\begin{equation}\label{angular1}
\Omega_{c}=\left(\frac{L}{E} \frac{f}{r^2}\right)\bigg| _{r_c}= \sqrt{\frac{f}{r^2}}\bigg| _{r_c};
\end{equation}

in the previous expressions we have incorporated the conditions for a circular orbit, $V (r_c)=0$ and $V'(r_c)=0$; these conditions amount, respectively, to

\begin{equation}\label{nullconds}
\frac{E^2}{L^2}= \frac{f }{r^2}\bigg| _{r_c}, \quad (2f - r f')\bigg| _{r_c}=0.
\end{equation}

In  \cite{Breton2016} were derived the expressions for the QNM in the case of a NLED Lagrangian $\mathcal{L} (F)$ that depends  only on the invariant $F$. 
At this point is worth to mention that in \cite{Konoplya2017} was analyzed the validity of the correspondence between the QNM in the eikonal approximation and the unstable null geodesics, finding that this correspondence does not hold for the Einstein-Lovelock theories, concretely in the case of the Einstein-Gauss-Bonnet BH. On the other hand in \cite{OSarbach2016} it was tested the convergence of the QNM, calculated numerically, to the eikonal approximation for a NLED  deviation from Maxwell theory. 
 In the following  subsection the effective potentials and the corresponding QNM expressions will be written for the EEH-BH, solution of the Einstein Eqs. coupled to the EH-Lagrangian $\mathcal{L}_{\rm EH} (F,G)$, Eq. (\ref{E-HLagrangian}).
\section{The QNM of the EEH--BH in the eikonal approximation}

The effective potential for a test photon with impact parameter $b= L/E$ is the neighborhood of the   EEH--BH, from $\dot{r}^{2} + V(r)=0$, is given by

\begin{equation}\label{Pot}
V_{i}(r)= E^2 \left(\frac{G^{e}_i}{G^{m}_i} \right)^2  \left(1 - b^2 \frac{G^{m}_{i} f}{G^{e}_i r^2}  \right),    
\end{equation} 
where  $G^{e}_{i}$ and $G^{m}_i$ are the electric and magnetic factors in the effective metrics; in the purely electric EEH--BH, $G^{m}_{i}=1$ and in the purely magnetic case $G^{e}_{i}=1$; in the RN case ($\mu=0$) $G^{e}_{i}=G^{m}_{i}=1$.
The conditions $V_{i}(r)=0$ and $V_{i} ^{'}(r)=0$ render the equations to determine the radius of the circular orbits, $r_{i_c}$ 

\begin{equation}
\left(\frac{f'}{f} - \frac{2}{r}\right) \bigg |_{r_{i_c}}= \left( \frac{G^{e}_i}{G^{m}_i} \right)^{'} \left( \frac{ G^{m}_{i}}{G^{e}_{i}} \right) \bigg |_{r_{i_c}}.
\label{rcEH}
\end{equation}
The corresponding solution should be greater than the horizon radius, $r_{i_c}>r_+$. The additional condition that defines the sphere of unstable null photon geodesics is $V^{''} (r_c) < 0$.
This radius is also known as the radius of the photosphere, that for a Schwarzschild BH is $r_c=3M$.

The Lyapunov exponent in the EEH--BH case is given by
\begin{equation}
\lambda^2_{i}= \frac{-V''_{i}}{2 \dot{t}^2}=\frac{fr^{2}}{2}\left( \frac{f}{r^{2}} \left( \frac{G^{e}_i}{G^{m}_i} \right)^{''} \left( \frac{ G^{m}_{i}}{G^{e}_{i}} \right) - \left(\frac{f}{r^{2}} \right)^{''} \right) \bigg  |_{r_{i_c}}, \quad i=1,2.   
\label{expLyapunovi}
\end{equation}
While the angular velocity (\ref{angular1}), is given by

\begin{equation}\label{angulari}
\Omega_{c}^{(i)}= \sqrt{\frac{G^{m}_{i}}{G^{e}_{i}} \frac{f}{r^2}}  \bigg|_{r_{i_c}} \quad i=1,2.
\end{equation}

Equations (\ref{expLyapunovi}) and (\ref{angulari}), the EEH-BH versions of  (\ref{Lyapunov}) and (\ref{angular1}),  determine the quasinormal frequencies, imaginary and real parts, respectively,  for  nonlinear electromagnetic EEH--BH in the  eikonal approximation. In what follows the EEH--BH  QNM are given explicitly for the electric case and are compared with those corresponding to massless test particles coming from the  linear electromagnetic Reissner-Nordstrom (RN) BH, that for completeness
we include in the next subsection.

\subsection{The QNM of the Reissner-Nordstrom black hole}

The RN is the SSS solution to the Einstein-Maxwell Eqs., i.e.  $\mathcal{L}=-F/4$, $\mathcal{L}_{F}=-1/4$ and $\mathcal{L}_{FF}=0$ .
In the eikonal or geometric-optics limit, the QNM are given by (\ref{QNM}) with 
$\lambda$ and $\Omega_{c}$ calculated as in   (\ref{Lyapunov}) and (\ref{angular1}), respectively. For the RN black hole, the metric function in the line element  (\ref{metric}) is given by Eq. (\ref{gtt}) with $\mu=0$. The null circular orbit radius $r_c$  is calculated from (\ref{nullconds}), that in the RN case amount to the quadratic polynomial and root, respectively,

\begin{equation}\label{rcRN}
r^2-3r+2Q^2=0, \quad r_c=  \frac{3}{2} (1 + \sqrt{1 - 8 Q^2/9}).
\end{equation}
 The functions $\lambda$ and $\Omega_c$ are given by  (\ref{Lyapunov}) and (\ref{angular1}) as

\begin{equation}
M^{2}\lambda^2= \frac{1}{r_c^6}[r^2-2Q^2][Q^2+r_c^2-2r_c], \quad M^{2}\Omega_c^2=\frac{1}{r_c^4}[Q^2+r_c^2-2r_c],
\end{equation}

that substituting $r_c$ from (\ref{rcRN}) gives

\begin{equation}
M^{2}\lambda^2= \frac{4 \sqrt{1-8Q^2/9}(1+3 \sqrt{1-8Q^2/9})}{3^3(1+ \sqrt{1-8Q^2/9})^4}, \quad M^{2}\Omega_c^2=\frac{2(1+ 3\sqrt{1-8Q^2/9})}{27(1+ \sqrt{1-8Q^2/9})^3}.
\end{equation}

These are the expressions for $\lambda$ and $\Omega$ for RN-BH;  we will compare them with the ones for EEH--BH. In the RN case analytic solutions can be found all the way through, however in the EEH case no analytic $r_c$ solutions were determined.

\subsection{ Electric EEH--BH QNM in the eikonal approximation}

For the analysis of the EEH-BH  QNMs, we express the black hole charge, the parameter $\mu$ and the radial distance in dimensionless form: $r \to r/M$, $Q \to Q/M$ and $\mu \to \mu /M^{2}$.  Only the fundamental frequency $n = 0$ is considered, that is, the least damped mode. We shall present the explicit results for the electric case, i.e. in this subsection we consider $G^{m}_{i}=1$ 

The circular null orbit radii $r_{i_c}$ are obtained from (\ref{rcEH}), that in the electric case  are the roots of 

\begin{equation}\label{roots}
5 r_{i_c}^4 [2 Q_e^2+ r_{i_c}( r_{i_c}-3)]- \mu Q_e^2 [Q_e^2+5 a_i  r_{i_c}( r_{i_c}-1)] =0,
\end{equation}
where $a_1=7$ and $a_2=4$. The analysis of the roots $r_{1_c}$ and $r_{2_c}$ of the Eqs. (\ref{roots})  is performed numerically in the ranges of $Q_e$ and $\mu$ where the EEH solution (\ref{gtt}) represents a black hole. 

\begin{figure}[h]
\begin{center}
\includegraphics [width =0.5 \textwidth ]{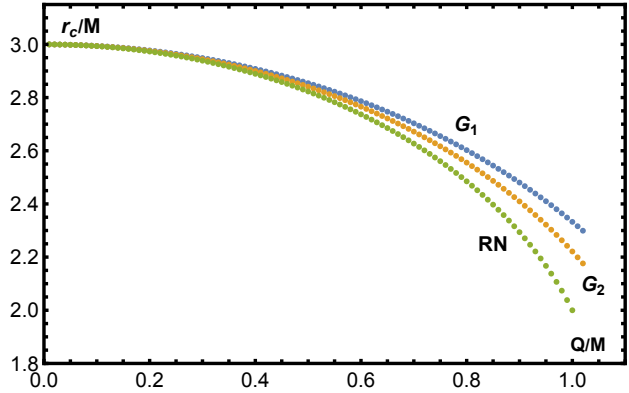}
\end{center}
\caption{The behavior of the  null circular orbit radii as a function of  $Q/M$ is shown for the electric EEH--BH and RN BH;
 the relative magnitudes of the radii are $r_{c}^{RN}< r_{2_{c}}^{EEH} < r_{1_{c}}^{EEH} < r_{c}^{S}$. The EH parameter $\mu$ is fixed to $\mu = 0.3$}
\label{Fig1.1}
\end{figure}

In Fig (\ref{Fig1.1}) is shown the tendency of the roots of the Eq. (\ref{roots}) in the case of EEH--BH and Eq. (\ref{rcRN}) for RN, as a function of $Q_e/M$. The null orbit radii $r_{i_c}$ for EEH--BH approach the corresponding to RN for small $Q_e$; $Q_e=0$ corresponds to the Schwarzschild photosphere, $r_{c}^{S}=3M$. Increasing the charge makes $r_{c}$ to decrease. In any case the relative magnitudes of the radii are $r_{c}^{RN}< r_{2_{c}}^{EEH} < r_{1_{c}}^{EEH} < r_{c}^{S}$.

The Lyapunov exponents $\lambda_i$ of the  EEH--BH are given by (\ref{expLyapunovi}),

\begin{equation}
M^{2}\lambda_{i}^{2}= \frac{1}{r^6} [r^2-2Q_e^2][Q_e^2+r(r-2)] - \frac{\mu Q^2}{20 r^{10}} \left[ Q_e^2 (19r^2-40 r+ 22 Q_e^2) +40 a_i [Q_e^2+r(r-2)]^2 \right] \bigg |_{r_{i_c}},
\label{eeh-lambda}
\end{equation}
where $a_1=7$ and $a_2=4$. In the previous  expressions we kept the terms up to $\mathcal{O}(\mu)$, to be consistent with the EH theory.


While the orbital angular velocities $\Omega_{c}^{(i)}$ (\ref{angulari}) are given by;

\begin{equation}
M^{2}(\Omega_{c}^{(i)})^{2}= \left( \frac{1}{r^4} [Q_e^2+r (r-2)] \left( 1 - 2 \mu \frac{a_i Q_e^2}{r^4}  \right) - \mu \frac{Q_e^2}{20r^8} \right)   \bigg |_{r_{i_c}},
\label{eeh-omega}
\end{equation}
where $a_1=7$ and $a_2=4$ and  we kept the terms up to $\mathcal{O}(\mu)$ as well.

\begin{figure}[h]
\begin{center}
\includegraphics [width =0.5 \textwidth ]{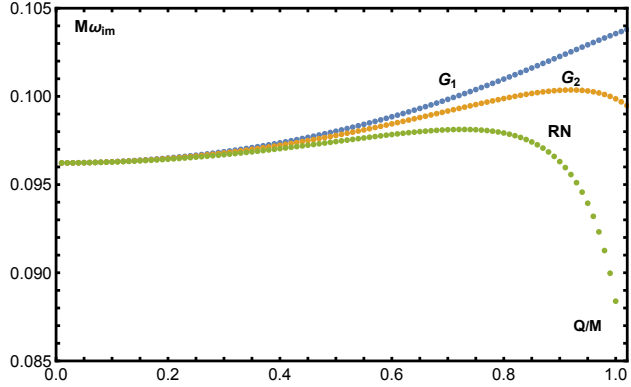}
\end{center}
\caption{The behavior of the imaginary part of the QNM,  $\omega_{\rm im}$, as a function of  $Q/M$ is shown for the electric EEH--BH and RN--BH;the effect of the EH parameter $\mu$ is of enhancing $\omega_{\rm im}$, implying this a shorter time for the damping of the perturbations. The rest of the parameters are fixed to $\mu = 0.3$ and $n=0$.}
\label{Fig1}
\end{figure}

In Fig. \ref{Fig1} the behavior of the imaginary QNM frequencies $\omega_{\rm im}$ of the EEH--BH for two different values of $G^{e}_{i}$ are shown; also are compared with the corresponding to RN BH. The imaginary part of the QNM $\omega_{\rm im}$, increases as $Q_e$ augments and after a maximum then decreases. For the RN-BH to have horizons the charge is constrained to $Q^2 < M^2$,
the EEH--BH does not have this constraint as a consequence of the screening of the charge;  $\omega_{\rm im}$ increases as $Q_e$ augments; for $G^{e}_{2}$ the curve presents a maximum that can fade off depending on the value of $\mu$.
The imaginary QNM $\omega_{\rm im}$  coming from EEH-BH considering both factors $G^{e}_{1}$ and $G^{e}_{2}$ are compared in Fig. 2.  for different values of the parameter $\mu$. For fixed $Q_e$ the QNM frequencies increase when $\mu$ increases and $\omega_{\rm im 1} > \omega_{\rm im 2}$.
\begin{figure}[h]
\begin{center}
\includegraphics [width =0.5 \textwidth ]{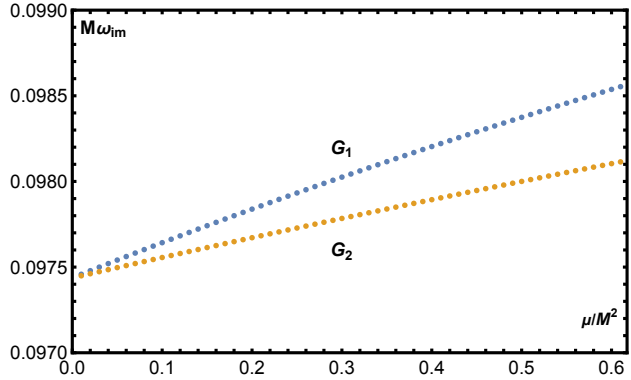}
\end{center}
\caption{The behavior of the imaginary part of the QNM,  $\omega_{\rm im}$, as a function of  $\mu$ is shown for the electric EEH--BH; the effect of the EH parameter $\mu$ is of enhancing $\omega_{\rm im}$ more for $G^{e}_{1}$, $\omega_{\rm im 1} > \omega_{\rm im 2}$;  implying this a shorter time for the damping of the perturbations; the intersection with the axis ($\mu=0$) is the value of $\omega_{\rm im }$ for the RN--BH. The dependence of $ \omega_{\rm im}$ on $\mu$ is linear. The charge parameter is  fixed to $Q = 0.5$ and $n=0$ }
\label{Fig2}
\end{figure}
For the analysis of the real part of QNM, we consider $\omega_{r} / l \to \omega_{r}$.  In Fig. \ref{Fig3}  the behavior of the real QNM frequencies $\omega_{r}$ of the EEH--BH  for different values of $Q_e$ is shown and compared with RN--BH. $\omega_{r}$  approaches the corresponding to  RN  as  $Q_e$ decreases. Increasing the charge $Q_e$ the real QNM $\omega_{r}$ increases. 
\begin{figure}[h]
\begin{center}
\includegraphics [width =0.5 \textwidth ]{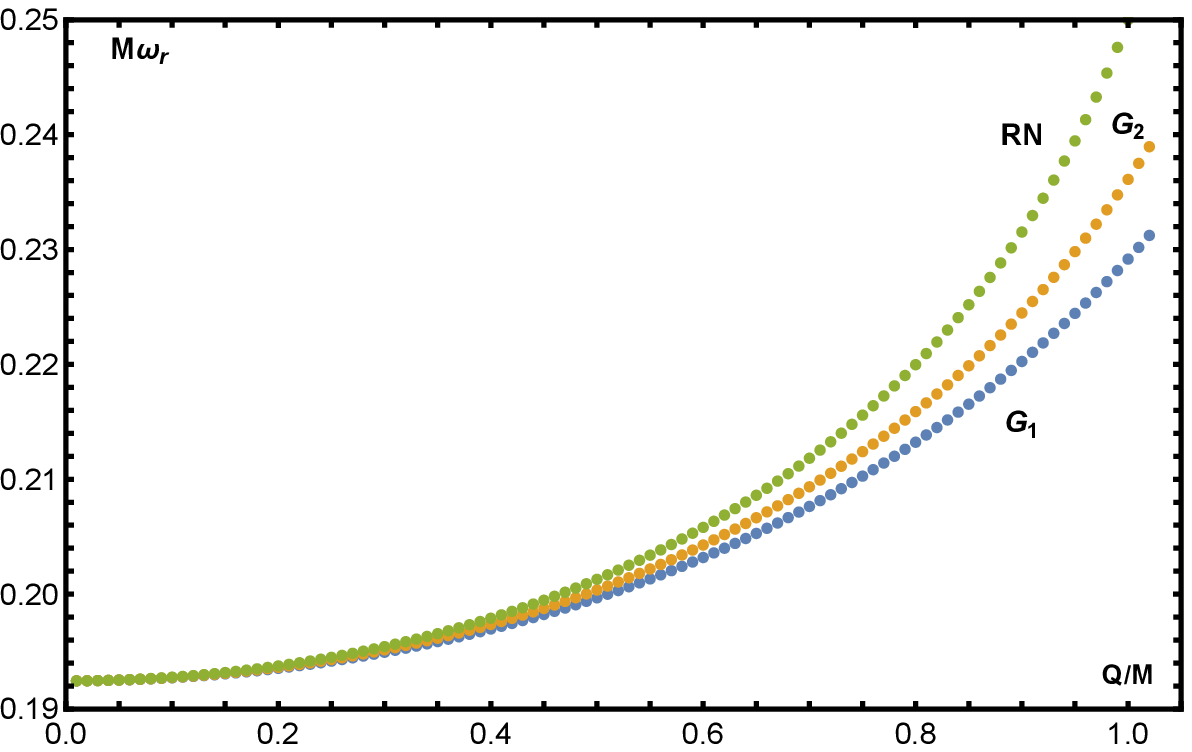}
\end{center}
\caption{The behavior of the real part of the QNM $\omega_{r}$ as a  functions of  $Q$  for the EEH--BH and RN--BH is shown; the EH parameter $\mu$ is fixed to $\mu = 0.3$}
\label{Fig3}
\end{figure}
For  fixed $Q_e$ the QNM frequencies $\omega_{r}$ decrease when the parameter $\mu$ increases. The real part of the frequencies coming from EEH--BH are compared in Fig. \ref{Fig4}. This corresponds to perturbations with slower oscillations.
\begin{figure}[h]
\begin{center}
\includegraphics [width =0.5 \textwidth ]{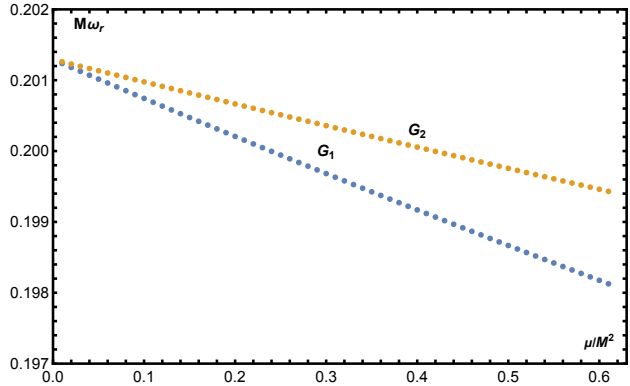}
\end{center}
\caption{The behavior of the real part of $\omega_{r}$ as a function of  $\mu$ is shown for the EEH--BH; $\omega_{r 1} < \omega_{r 2}$, both varying linearly with $\mu$. The charge is fixed to $Q = 0.5$}
\label{Fig4}
\end{figure}
Therefore corresponding to the two possible trajectories for light rays in the vicinity of the EEH-BH, there will appear two QNM. In principle, if light were polarized  it would be a distinction. In the case of unpolarized light impinging on the EEH-BH, the trajectory followed would be obtained by taking an average over  the two polarization modes
\cite{Novello2000b}, \cite{Bialynicka1970}. At the present state of the observational facilities, most likely the difference between the two trajectories would be still unnoticeable.

\section{Conclusions}

In the Euler--Heisenberg nonlinear electrodynamics the birefringence effect occurs, in such a way that there are two effective metrics whose null geodesics are the light trajectories, and following one or the other depends on light polarization. Effective metrics are obtained from the background metric, in this case the EEH--BH metric,  but taking into account the NLEM  effects.
We have determined the two effective metrics followed by photons for both, the electric and the magnetic EEH-BH.

As an application of birefringence we have calculated the QNM frequencies of the EEH-BH in the geometric optical approximation. QNM frequencies were calculated from the two possible null unstable geodesics of the effective metrics.
From the expressions of the real and imaginary parts of the QNM frequencies and Eqs.  (\ref{expLyapunovi}) and (\ref{angulari}), it is clear that the NLEM effects will modify QNM frequencies enhancing the imaginary part $\omega_{\rm im}$  and suppressing the real one $\omega_r$.
This comes from the modifying factors, electric $G^{e}_{i}$, or magnetic $G^{m}_{i}$, $i=1,2$ in the effective metrics. 

The  comparison is done with the QNM frequencies of  the  Maxwell linear counterpart, the RN--BH. As can be  observed in Figs. \ref{Fig1} and \ref{Fig3}, the NLEM effect when varying the electric charge $Q_e$ is of suppressing the real part while increasing the imaginary one, i.e. oscillation periods are shorter and relaxation occurs faster. On the other hand, the effect of the EH parameter $\mu$ is of screening the electric charge, that renders a more Schwarzschild-like behavior in general. In the magnetic case the corresponding QNM, that we do not explore in detail,  follow the same tendencies than the electric ones, this can be asserted from the expressions for the Lyapunov exponent and the angular velocity, Eqs. (\ref{expLyapunovi}) and (\ref{angulari}),  in that the magnetic factor $G^{m}_{i}$ appears inversely than the electric $G^{e}_{i}$ one,
but $G^{e}_{i} > 1$ while $G^{m}_{i} < 1, \quad i=1,2. $
Also we plot the QNM frequencies varying  $\mu$; it is interesting to note that when  $\mu$ increases, the imaginary part of the QNM continues  increasing  pointing out to a shorter time for restoring the unperturbed state of the EEH--BH as compared to the linear case RN--BH.
The opposite tendency occurs with the real part of the QNM frequencies, $\omega_{r}$, that are suppressed  as $\mu$ increases.
Finally we can mention that a thorough analysis is needed to know how the modifications introduced by NLEM effects influence the BH  stability; steps in this direction are given in \cite{OSarbach2016}. 
\vspace{0.5cm}

\textbf{Acknowledgments}: 
N. B. acknowledges partial financial support from CONACYT-Mexico through the project No. 284489. The authors acknowledge  financial support from SNI-CONACYT, Mexico.


\begin{thebibliography}{9}

\bibitem{Gold1968}
T. Gold, {\it Rotating Neutron Stars as the Origin of the Pulsating Radio Sources},
Nature, {\bf 218}, 731-732 (1968).

\bibitem{Baring2008}
M. G. Baring, {\it Photon Splitting and Pair Conversion in Strong Magnetic Fields},
AIP Conf. Proc. {\bf 1051}:53-64 (2008), [arXiv:  0804.0832]

\bibitem{Bialynicka1970}
Z. Bialynicka-Birula, I. Bialynicki-Birula,
{\it Nonlinear Effects in Quantum Electrodynamics. Photon Propagation and Photon Splitting in an External Field}, Phys. Rev. D {\bf 2} 2341-2345 (1970).

\bibitem{Brezin1971}
E. Brezin, C. Itzykson, {\it Polarization Phenomena in Vacuum Nonlinear Electrodynamics},
Phys. Rev. D {\bf 3} 618-621 (1971)

\bibitem{deMelo2015}
C. de Melo, L. Medeiros, P. Pompeia, {\it Causal structure and birefringence in nonlinear electrodynamics}, Mod. Phys. Lett. A {\bf 30} (06)1550025 (2015)

\bibitem{Luiten2004}
A. N. Luiten, J. C. Petersen, {\it Detection of vacuum birefringence with intense laser pulses}, Phys Lett. A {\bf 330} (6), 429-434 (2004)

\bibitem{Gies2009}
H. Gies, {\it Strong laser fields as a probe for fundamental physics,} Eur. Phys. J. {\bf D55},
311-317 (2009) [arXiv:0812.0668].

\bibitem{Karbstein2020}
F. Karbstein,
{\it Probing vacuum polarization effects with high-intensity lasers},
Particles, {\bf 3}(1), 39-61 (2020)

\bibitem{Brodin2001}
G. Brodin, M.  Marklund  and L. Stenflo,  {\it Detection of QED vacuum nonlinearities in Maxwell's equations by the use of waveguides}, Phys. Rev. Lett. {\bf 87} 171801 (2001).

\bibitem{EK1935}
H. Euler and B. Kockel, {\it The scattering of light by light in the Dirac theory},
Naturwiss. {\bf 23}, 246 (1935).

\bibitem{EH1936}
W. Heisenberg and H. Euler, {\it Folgerungen aus der diracschen theorie des positrons}. Zeitschrift Für Physik,  {\bf 98} (11-12), 714-732 (1936); English translation: {\it Consequences of Dirac's Theory of Positrons,} arXiv: physics/0605038

\bibitem{Ferrari1984}V. Ferrari and B. Mashhoon, New approach to the quasinormal modes of a black hole , Phys. Rev. D, {\bf 30}   295-304 (1984)
 
\bibitem{Cardoso2008}
V. Cardoso, A. S. Miranda, E. Berti, H. Witek, T. V. Zanchin,
{\it Geodesic stability, Lyapunov exponents and quasinormal modes}, 
Phys. Rev.{\bf D79}, 064016 (2009); arXiv: hep-th/0812.1806; 
 
\bibitem{Pleban}
S. Alarcon Gutierrez, A. L.  Dudley and J. F. Plebański,  {\it Signals and dicontinuities in general
relativistic non-linear electrodynamics} J. Math. Phys. {\bf 22} 2835–48 (1981)

\bibitem{Novello2000}
V. A. De Lorenci, R. Klippert, M. Novello,  J. M. Salim, 
{\it Light propagation in nonlinear electrodynamics},  Phys. Lett. B {\bf 482} 134-140 (2000)

\bibitem{Remo2013} R. Ruffini, Yuan-Bin Wu, and She-Sheng Xue: {\sl Einstein-Euler-Heisenberg theory and
charged black holes}, {\em Phys. Rev.} {\bf D88} (2013) 085004.

\bibitem{Amaro2020} D. Amaro and A. Mac\'{\i}as: {\sl Geodesic Structure of the Euler--Heisenberg Static Black Hole}, 
{\em Phys. Rev.} {\bf D102} (2020) 104054.

\bibitem{Mann}
S. Gunasekaran, D. Kubizn\'ak and R. B. Mann, {\it Extended phase space
thermodynamics for charged and rotating black holes and Born-Infeld vacuum
polarization}, J. High Energ. Phys. {\bf 2012}, 110 (2012). 

\bibitem{Kruglov2010}
S. I. Kruglov, {\it On generalized Born–Infeld electrodynamics},
J. Phys. A: Math. Theor. {\bf 43} (2010) 375402 (8pp).


\bibitem{Toshmatov2017}
B. Toshmatov, Z. Stuchl\'\i{}k and B. Ahmedov, {\it Rotating black hole solutions with quintessential energy}, Eur. Phys. J. Plus. {\bf 132}, 98 (2017).

\bibitem{Yajima2001}
H. Yajima and T. Tamaki, {\it Black hole solutions in Euler-Heisenberg theory}, Phys. Rev. D {\bf 63} 064007 (2001).

\bibitem{Rubiera2019}
M.  Guerrero, D.  Rubiera-Garcia. {\it Nonsingular black holes in nonlinear
gravity coupled to Euler-Heisenberg electrodynamics}. Physics Letters B
{\bf 788}, 446-452  (2019).

\bibitem{Kruglov2017}
S. I. Kruglov, {\it Remarks on Heisenberg-Euler-type electrodynamics},
Mod. Phys. Lett. {\bf A 32} 1750092 (2017).

\bibitem{Macias2019}
N. Breton, C. L\"ammerzah and A. Mac\'{\i}as,
{\it Rotating Black Holes in the Einstein--Euler--Heisenberg theory},
Class. and Quantum Grav. {\bf 36 }, 235022 (22pp) (2019).

\bibitem{Novello2000b}
M. Novello, V. A. De Lorenci,  J. M. Salim  and R. Klippert, {\it Geometrical aspects of light propagation in nonlinear electrodynamics},  Phys. Rev. D {\bf 61} 045001 (2000)

\bibitem{Liberati-Sonego-Visser}
S. Liberati, S. Sonego, M. Visser, {\it Scharnhorst effect at oblique incidence},  
Phys. Rev. D {\bf 63} 085003 (2001)

\bibitem{Obukov2002}
Y. N. Obukhov and G. F. Rubilar, 
{\it  Fresnel analysis of wave propagation in nonlinear electrodynamics},
Phys. Rev. D {\bf 66}, 024042 (2002)

\bibitem{Goulart2009}
E. Goulart  and S. E. Perez Bergliaffa,  {\it A classification of the effective metric in nonlinear electrodynamics}, Classical and Quantum Gravity, {\bf 26}, 13, 135015 (2009).

\bibitem{Kruglov2015}
S. I. Kruglov, {\it Nonlinear electrodynamics with birefringence},
Phys.Lett. A {\bf 379} 623-625 (2015)

\bibitem{Breton2016}
N. Breton, L. A. Lopez,
{\it Quasinormal modes of nonlinear electromagnetic black holes
from unstable null geodesics}
Phys. Rev. D {\bf 94} 104008 (2016)

\bibitem{Konoplya2017}
R.A. Konoplya, Z. Stuchlík, {\it Are eikonal quasinormal modes linked to the unstable circular null geodesics? }
Phys.Lett. B {\bf 771} 597-602 (2017)

\bibitem{OSarbach2016}
E. Chaverra, J. C. Degollado, C. Moreno, and O. Sarbach,
{\it Black holes in nonlinear electrodynamics: quasinormal
spectra and parity splitting}, Phys. Rev. D {\bf 93}, 123013
(2016).

\end{thebibliography}
\end{document}